\documentclass[%
 reprint,
nofootinbib,
 amsmath,amssymb,
 aps,
]{revtex4-2}
\usepackage{pdfcomment}
\usepackage[colorinlistoftodos]{todonotes}
\usepackage{graphicx}
\usepackage{dcolumn}
\usepackage{bm}

\usepackage{amsmath}
\usepackage{subcaption}
\usepackage{comment}
\usepackage[justification=raggedright]{caption}

\graphicspath{{graphs/}}

\begin{document}

\preprint{APS/123-QED}

\title{Universal Waveforms for Extreme Mass-Ratio Inspiral}

\author{Ilan Strusberg}
 \email{ilan.strusberg@mail.huji.ac.il}
\author{Barak Rom}%
\author{Re'em Sari}
\affiliation{%
 Racah Institute of Physics, The Hebrew University of Jerusalem, 9190401, Israel
}%

\date{\today}

\begin{abstract}
We engage with the challenge of calculating the waveforms of gravitational waves emitted by spinless binary black hole merger in extreme mass-ratio limit.
We model the stellar-mass black hole as a test-particle, initially on a circular orbit, that undergoes adiabatic inspiral until it reaches the innermost stable circular orbit (ISCO), after which it follows a geodesic trajectory. We compute the gravitational waveforms emitted during both phases—before and after the ISCO crossing—and demonstrate how to accurately connect them. While the waveforms are calculated adiabatically up to the ISCO, the associated phase error near the ISCO scales as $\nu^{1/5}$ and remains below one radian for sufficiently small mass-ratios $\nu$. 
Our complete waveform is universal in the sense that all computationally expensive calculations are performed once, and its application to any binary merger can be obtained by appropriately re-scaling time, phase, and amplitude. 
We compare our results with existing models in the literature and show that our complete waveforms are accurate enough all the way from separations that are an order of one gravitational radii outside the ISCO, to the merger.
\end{abstract}

\maketitle


\section{Introduction}
\label{Intruduction}
Extreme Mass-Ratio Inspirals (EMRIs) occur when a stellar-mass black hole (sBH), of mass $m/M_\odot\in[1,10^2]$, forms a binary system with a supermassive black hole (SMBH) of mass $M/M_\odot\in[10^5,10^9]$. As the system evolves, gravitational wave (GW) emission leads to energy loss, causing the smaller object to spiral inward toward the SMBH. These events generate GWs in the mHz to Hz frequency range and are expected to be a key source for the Laser Interferometer Space Antenna (LISA) \cite{lisa_website, esa_lisa, LISAReqDoc}. Analyzing LISA data will be highly challenging and will require the ability to generate waveform templates with radian-level accuracy in a computationally efficient manner \cite{Babak_2017, lisa_consortium_waveform_working_group_waveform_2023}.

The inspiral of the compact object is driven by the gravitational self-force (GSF), which arises from the small perturbation that the sBH induces in the background  metric of the SMBH \cite{barack_self_force_2018, barack_gravitational_2009, Pound2021}. This inspiral unfolds over a radiation-reaction timescale $t_{rad}\sim \left(M/m\right)t_{dyn}$, which is significantly longer than the dynamical timescale $t_{dyn}\sim GM/c^3$. This leads to the \emph{adiabatic approximation}, where the compact object gradually transitions from one geodesic orbit in the background metric to another due to orbit-averaged GW emission, computed to leading order in the mass-ratio. The adiabatic approximation has been widely studied in the literature, and numerous models have used it to estimate the GW signals emitted by EMRIs \cite{Hughes_2021, Hughes2000, Mino_2003, drasco_gravitational_2006, Isoyama_2022,Taracchini_2014, Kennefick_1996, Detweiler1978, Samuel_2000, Gralla2016, Nasipak_2024}. One prominent model is the Fast-EMRI-Waveforms (FEW) \cite{chua_rapid_2021, Katz_2021}, regarded as the gold standard for adiabatic waveforms. Several models have gone beyond the adiabatic approximation, incorporating \emph{post-adiabatic} effects into their waveform templates \cite{wardell_gravitational_2023, mathews2025postadiabaticwaveformgenerationframeworkasymmetric}. Numerical Relativity (NR) was the first to generate waveforms for binary systems in the comparable to intermediate mass-ratio regime \cite{pretorius_evolution_2005, baker_gravitational-wave_2006, Campanelli_2006}. The Effective One Body (EOB) approach \cite{Buonanno_1999,nagar_time-domain_2018, pompili2023layingfoundationeffectiveonebodywaveform}, which provides a robust framework for modeling EMRI waveforms, has already produced waveforms that include post-adiabatic effects for binary systems with aligned-spin components \cite{albertini2024comparingeffectiveonebodymathissonpapapetroudixonresults, Albertini_2024}. It has also been utilized in the data analysis of the LVK collaboration \cite{pompili2023layingfoundationeffectiveonebodywaveform}.

In this paper, we study the GW emission during the merger of a binary black hole (BH) system in the extreme mass-ratio limit. We focus on circular EMRIs of non-spinning BHs.
In this case, the evolution of the system can be divided into two main stages: an \emph{adiabatic inspiral} followed by a \emph{rapid plunge}. During the first stage, the sBH slowly descends from one circular orbit to another, while in the second stage, its trajectory approaches a geodesic \cite{bernuzzi_binary_2010, Sundararajan_2010, PhysRevD.90.084025,buonanno_transition_2000, Sperhake_2008, Baker_2001, Price_2016}, known as the geodesic universal infall (GUI) trajectory \cite{rom_extreme_2022}.

We present a simple method to produce a universal waveform that accurately captures the final stages of the merger,
from a few gravitational radii outside the ISCO onward, which can be easily rescaled for any given mass-ratio. Therefore, this procedure is computationally efficient and may be useful for template-based searches. 

Our method sharply connects the adiabatic evolution outside the ISCO and the plunging phase inside it. We show that such a sharp connection, with no consideration for a transition regime \cite{Ori_2000}, is sufficiently accurate for waveform generation in the extreme mass-ratio limit. In this paper, we focus on calculating the dominant mode of the waveform (corresponding to the multipolar indices $l=m=2$). However, our method can be simply generalized to compute other modes as well.
Our results are implemented in the \emph{Universal Waveforms} code, available at \cite{UniversalWaveformsCode}.

In Section \ref{sec: main}, we discuss the two stages of the merger, the adiabatic inspiral and the geodesic plunge (Sections \ref{sec:AD} and \ref{sec:GUI}, respectively), and outline our method for constructing the complete merger waveform by connecting these two results (Section \ref{sec: Full}). 
In Section \ref{sec: results}, we evaluate the phase error accumulated in the adiabatic and GUI waveforms (Sections \ref{sec:err_AD} and \ref{sec:err_GUI}, respectively). In Section \ref{sec: Comparison}, we compare our complete waveform to the FEW \cite{chua_rapid_2021, Katz_2021} and the TEOBResumS \cite{nagar_time-domain_2018, Albertini_2024} waveforms. We summarize our results in Section \ref{sec: Conclusions}.

Throughout the paper we use geometric units, $G=c=1$, and measure distances and time in units of the SMBH mass $M$. 

\section{GW emission in the extreme mass-ratio limit}\label{sec: main}
We consider a sBH, with mass $m$, orbiting a SMBH, with mass $M\gg m$.
Assuming the sBH initially follows a circular orbit, we approximate its trajectory by treating it as a test-particle moving under the influence of the effective Schwarzschild potential
\begin{equation}
    V_{eff}(r,l)=\left(1-\frac{2}{r}\right)\left(1+\frac{l(\tau)^2}{r^2}\right),   
    \label{eq: Schwarzschild Effective Potential}
\end{equation}
where $l(\tau)$ is the angular momentum per unit mass, which decreases over time due to GW emission. 
Consequently, the equation of motion for the sBH, with respect to its proper time $\tau$, is
\begin{equation}
    \ddot{r}=-\frac{1}{2}\frac{\partial V_{eff}(r,l)}{\partial r}.
    \label{equation of motion of the test-particle}
\end{equation}

The orbital evolution, described by the above equation, can be decomposed into two regions: 
(I) Outside the ISCO, where the sBH roughly follows the minimum of the effective potential. In this region, the orbital evolution can be approximated as an adiabatic quasicircular inspiral, where the sBH slowly descends from one circular orbit to another  with a radial velocity determined by the orbit-averaged GW emission.
(II) Inside the ISCO, where the angular momentum loss becomes
negligible, so the sBH follows a plunging geodesic. Thus, its trajectory can be approximated by the GUI \cite{rom_extreme_2022}, as given by Eq. (\ref{equation of motion of the test-particle}), with a constant angular momentum and initial conditions corresponding to a circular orbit at the ISCO, $l_{ISCO}=\sqrt{12}$ and $R_{ISCO}=6$.

Note that in Eq. (\ref{equation of motion of the test-particle}), we neglect the conservative force, which is of order $O(\nu)$, as it does not contribute to the adiabatic order and its impact on the phase remains negligible near the ISCO compared to the dynamical effects \cite{Pound_2014,Barack_2009}.

We construct the complete merger waveform (Section \ref{sec: Full}) by sharply connecting the waveforms associated with the two orbital evolution phases, the adiabatic inspiral and the geodesic plunge.

The latter, which is formally infinitely long, is truncated at a finite time determined by the Ori-Thorne transition formalism \citep{Ori_2000}, as discussed below.

We thus obtain a universal waveform, in the sense that it can be simply adapted to any given symmetric mass-ratio $\nu \ll 1$. 
This waveform faithfully captures the final stages of the merger, spanning a timescale of $t\sim\nu^{-1}$, or equivalently, covering the region from separations that are an order of unity larger than the ISCO, to the merger.

\subsection{Outside the ISCO: the adiabatic approximation}\label{sec:AD}
The flux-driven adiabatic waveforms have been extensively studied in the literature \cite{Taracchini_2014, Hughes2000, Mino_2003, drasco_gravitational_2006, Isoyama_2022, Kennefick_1996, Detweiler1978, Samuel_2000, Gralla2016, Nasipak_2024}.
In this approximation, the sBH motion can be described as a transition from one circular orbit to another, since $\dot{r}/r\ll\Omega$, where r is the orbital radius, $\dot{r}$ is the radial velocity and $\Omega(r)=r^{-\frac{3}{2}}$ is the orbital frequency.

Therefore, the equations of motion are given by 
\begin{equation}
    \begin{aligned}
        \frac{dE}{d\Tilde{t}}&=-P_{cir}\left(r\right),
       \\
        \frac{d\Tilde{\phi}}{d\Tilde{t}}&=m\Omega\left(r\right),
        \label{eq:Adiabatic equations}
    \end{aligned}
\end{equation}
where $\Tilde{t}=\nu t$, $\Tilde{\phi}=\nu\phi$ are the normalized time and angular coordinate, respectively, and $E=\left(r-2\right)/\sqrt{r(r-3)}$ is the energy per unit mass on a circular orbit. $P_{cir}(r)$ is the GW luminosity per unit mass, emitted by a small mass orbiting on a circular orbit of radius $r$, divided by the mass-ratio $\nu$.
We calculate $P_{\rm cir}(r)$ numerically, up to first order in $\nu$, with high fractional accuracy of $10^{-12}$, as described in Appendix \ref{app: ODE}.
We then solve Eqs. (\ref{eq:Adiabatic equations}) for the orbital radius $R\left(\Tilde{t}\right)$ and the normalized angular coordinate $\Tilde{\phi}\left(\Tilde{t}\right)$
backward in time, i.e., from the ISCO outwards, with the initial conditions of a circular orbit at the ISCO:
$R(0)=6$ (or $E(0)=2\sqrt{2}/3$) and $\Tilde{\phi}(0)=0$.\\
\begin{figure}[h!]
    \centering
    \includegraphics[width=1\linewidth]{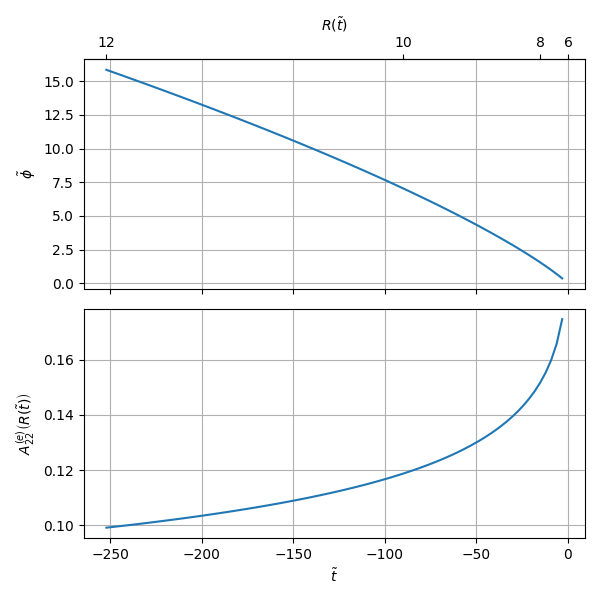}
    \caption{The normalized waveform of the
    adiabatic inspiral, 
    i.e., the solution of Eq. (\ref{eq:Adiabatic equations}). Top: normalized angular coordinate $\Tilde{\phi}$,
    bottom: normalized amplitude of the dominant mode of the adiabatic wave $A_{22}^{(e)}$.
    Both are given as a function of the normalized time $\Tilde{t}$, which is $0$ when the test-particle is at the ISCO, and the position of the sBH $R(\Tilde{t})$ (top axis).}
    \label{fig: Adiabatic Phase and Amplitude}
\end{figure}
The waveform of the GW detected by a distant observer during the adiabatic inspiral is given by 
 \begin{equation}
    \psi^{(\lambda)}_{\ell m}(t)=\nu A^{(\lambda)}_{\ell m}\left(R\left(\nu t\right)\right)\exp\left(im\Tilde{\phi}\left(\nu t\right)/\nu\right),
    \label{AD waveforms}
\end{equation}
where $\lambda$ is the parity, $(\ell, m)$ are the multipolar indices and $A_{\ell m}^{(\lambda)}(r)$ is the normalized amplitude of the GW emitted by a test-particle on a circular orbit of radius $r$, as discussed in Appendix \ref{app: ODE}.

To enhance the computational efficiency in our \emph{Universal Waveforms} code \cite{UniversalWaveformsCode}, we calculate $\Tilde{\phi}(\Tilde{t})$ and $A_{22}^{(e)}(R(\tilde{t}))$ on a grid with a timestep of $\Delta \Tilde{t}=2.5\times10^{-4}$, and evaluate them using cubic spline interpolation. This interpolation introduces a negligible phase error $\phi$, which remains below $10^{-14}/\nu$.

In Fig. (\ref{fig: Adiabatic Phase and Amplitude}), we show the normalized angular coordinate $\Tilde{\phi}\left(\Tilde{t}\right)$ and the normalized amplitude of the dominant mode $A_{22}^{(e)}\left(R\left(\Tilde{t}\right)\right)$ as a function of the normalized time $\Tilde{t}$. 

\subsection{GUI Waveform}\label{sec:GUI}
Shortly after the sBH crosses the ISCO, at a radial distance of order $\left(6-r\right)\sim\nu^{2/5}$ \citep{Ori_2000,buonanno_transition_2000}, its motion follows the GUI, as the energy dissipation (per unit mass) by GW from the ISCO crossing up to the merger is negligible (it scales as $\nu^{4/5} \ll 1)$.

The GUI waveform is therefore obtained by numerically solving the Regge-Wheeler-Zerilli \citep[RWZ;][]{ReggeWheeler,Zerilli} equation, with a source term
corresponding to a test-particle initially on a circular orbit at the ISCO \citep[for further details see][and references therein]{rom_extreme_2022}.\\
The waveform is given as a function of the Eddington–Finkelstein retarded time $u=t-r^*$, where $r^*=r+2\log\left(r/2-1\right)$ is the tortoise coordinate. 

We shift the waveform such that it peaks at $u=0$.
As in the adiabatic case, we evaluate the phase and amplitude of the GUI waveforms using a cubic spline interpolation (with a timestep of $\Delta u = 9\cdot10^{-3}$). The interpolation introduces a negligible error in both of them.
The results are subsequently incorporated into our \emph{Universal Waveforms} code \cite{UniversalWaveformsCode}.

\subsection{The complete Universal Waveform}\label{sec: Full} 
The complete merger waveform is obtained by connecting the adiabatic and the GUI waveforms, such that the total error in phase will be less than unity, as discussed in Section \ref{sec: results}. The waveform is given as a function of the retarded time $u$, and is shifted so that its peak will be at $u=0$.

We define the plunge time, $t_{\rm plunge}$, as the time interval between two events (I)  the angular momentum of the  test-particle is equal to the ISCO angular momentum $\ell_{\rm ISCO}=\sqrt{12}$, and (II) the emission of the  waveform peak, i.e. the test-particle is at $r_{peak}=3.764$ (as given in Appendix \ref{app: Delta t} ). Note that the first event is not equivalent to stating that $R=6$, as during the transition phase \cite{Ori_2000}, around the ISCO, the particle does not reside at the potential minimum.
Accordingly, the GUI waveform is truncated and connected to the adiabatic waveform at $u_{cut}$, which is determined by $t_{plunge}$ with an additional constant shift due to the signal travel time (for more details, see Appendix \ref{app: Delta t}).

To evaluate $t_{plunge}$, we describe the dynamics at the ISCO's vicinity using the Ori-Thorne formalism \citep{Ori_2000}.
Specifically, we consider a constant angular momentum loss rate such that
\begin{equation}
    \delta l=l(\tau)-\sqrt{12}=-\frac{\sqrt{2}P_{ISCO}}{\Omega_{ISCO}}\nu\tau,
    \label{angular momentum diff}
\end{equation}
where $\Omega_{ISCO}=\sqrt{6}/36 \cong 6.804\cdot10^{-2}$ and $P_{ISCO}=9.403\cdot10^{-4}$ are the angular frequency and GW luminosity at the ISCO, respectively.
In this region, the equation of motion (Eq. \ref{equation of motion of the test-particle}) can be expanded in the limit $x=\left(6-r\right)\ll1$, yielding
\begin{equation}
\begin{split}
    \ddot{x}&=-\frac{x^2}{6^{4}}\left(1+\frac{2}{3}x+O\left(x^2\right)\right)+O\left(\delta l\right).
    \label{expansion of equation of motion}
\end{split}
\end{equation}

The first-order term, $\ddot{x}\propto x^2$, leads to a plunge time that scales as $t\propto x^{-1/2}$. In the transition region $x\propto\nu^{2/5}$ and therefore $t\simeq101.6\nu^{-1/5}$, where the coefficient is obtained by numerically integrating the above equation of motion \citep[for further details see][]{Ori_2000}. 
The first-order correction to the acceleration, given by the $2x/3$ term inside the parentheses in Eq. (\ref{expansion of equation of motion}), leads, along times of order $t\propto x^{-1/2}$, to a constant, $\nu$-independent correction to $t_{\rm plunge}$.
The next order term in $t_{\rm plunge}$ scales as $\nu^{1/5}$ and is required for accurately determining  $u_{cut}$ at higher mass-ratio values ($\nu\gtrsim10^{-3}$). Our goal is to estimate $u_{cut}$ with  accuracy of order unity, as larger errors would lead to phase shift that will be accumulated on a timescale shorter than $\nu^{-1}$ during the adiabatic phase.
Thus, $u_{cut}$ is given by
\begin{equation}
\begin{aligned}
    u_{cut}=-101.6\nu^{-\frac{1}{5}}+108.8-A\nu^\frac{1}{5}.
    \label{GUI cutting time}
\end{aligned}
\end{equation}
\begin{figure}[h!]
    \centering
    \includegraphics[width=\linewidth]{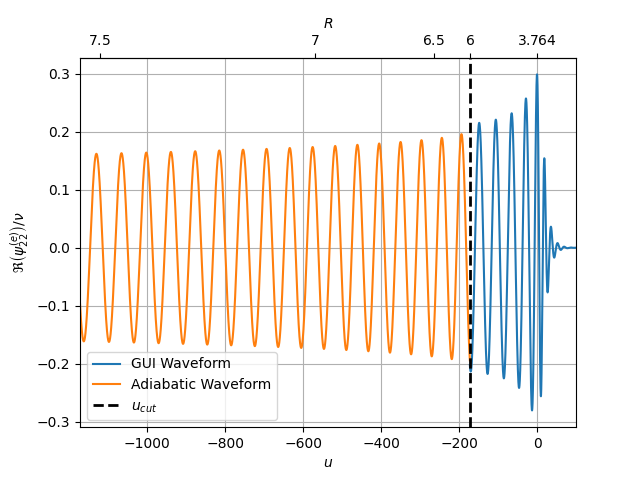}
    \caption{
    The complete EMRI waveform, for a mass-ratio of $\nu=10^{-2}$. We present the real part of the dominant mode, $\psi_{22}^{(e)}$, as a function of the retarded time, $u=t-r^*$, and the position of the test-particle (top axis). The waveform consists of an adiabatic part (in blue), representing the quasi-circular inspiral, and a GUI part (in orange), corresponding to the final geodesic plunge. The black dashed line located at $u_{cut}\approx-172$ (Eq. \ref{GUI cutting time}) is the point of transition between the adiabatic waveform and the GUI waveform.
    } 
    \label{fig: complete wave}
\end{figure}
Note that $u_{cut}<0$ since the adiabatic phase precedes the plunge phase. In Appendix \ref{app: Delta t} we derive the first two numerical coefficients in the equation above. 
The value of the last term in Eq. (\ref{GUI cutting time}), $A\approx65$, is determined numerically by solving the full dynamics near the ISCO (Eq. \ref{equation of motion of the test-particle}), with a constant angular momentum loss rate (Eq. \ref{angular momentum diff}). We determine $A$ with an accuracy of $O(1)$. This is sufficient  since it is the coefficient of $\nu^{1/5}<1$, and $u_{cut}$ needs to be determined to an accuracy of order unity.
Figure \ref{fig: complete wave} presents an example of the complete waveform, comprising both the adiabatic part and the GUI part, for a mass-ratio of $\nu=10^{-2}$, resulting in $u_{cut}\approx-172$. 

\section{Analysis and Comparison}\label{sec: results}
Due to our approximate treatment, the waveform, as presented in Fig. (\ref{fig: complete wave}), accumulates an error in phase.
At the vicinity of the ISCO, the error arises from deviations of the GUI trajectory and the adiabatic inspiral from the solution of the full dynamics (Eq. \ref{equation of motion of the test-particle}). 
At larger distances, the error stems from the higher-order corrections to the adiabatic approximation,  limiting the validity of our waveform up to separations of order unity beyond the ISCO.

\subsection{Phase Error in The GUI Waveform} \label{sec:err_GUI}
In the transition region, the test-particle's trajectory, as determined by Eq. (\ref{equation of motion of the test-particle}), differs from the GUI trajectory by $\Delta x_{GUI}\sim\nu^{2/5}$, for a duration of $\Delta t\sim\nu^{-1/5}$ \cite{Ori_2000}. 
Therefore, the GUI trajectory introduces an error in the angular frequency that scales as $\delta\Omega_{GUI}\propto\Delta x_{GUI}\sim\nu^{2/5}$,
which results in a phase shift of
\begin{equation}
    \Delta\phi_{GUI}\approx\delta\Omega_{GUI}\Delta t\sim\nu^\frac{1}{5}.
\end{equation}

Note that in the GUI trajectory we neglect the change in the particle's angular momentum. While this is an additional source of phase error, its contribution is negligible in the extreme mass-ratio limit, since the deviation in angular momentum per unit mass scales as $\nu^{4/5}$ \citep{Ori_2000,buonanno_transition_2000,rom_extreme_2022}, resulting in a phase error of order $\sim\nu^{3/5}$.
To quantitatively estimate the phase error, we compute the phase accumulated along the GUI trajectory, $\phi_{plunge}$, over our effective plunge time, $t_{\rm plunge}$ (given in Appendix \ref{app: Delta t})
\begin{equation}
    2\phi_{\rm plunge}=13.82\nu^{-\frac{1}{5}}-8.989+3.6\nu^\frac{1}{5}.
    \label{eq: phi plunge}
\end{equation}
Where, along the GUI, $t(r)$ and $\phi(r)$ are determined analytically (see Appendix \ref{app: Delta t}). 

We compare Eq. (\ref{eq: phi plunge}) to the total phase a test-particle accumulates from the ISCO's crossing until it reaches $r_{peak}=3.764$ using the solution of the full dynamics near the ISCO (Eqs. \ref{equation of motion of the test-particle} and \ref{angular momentum diff}).
From this comparison, we reproduce the expected scaling, $\Delta\phi \propto \nu^{1/5}$, and determine the numerical coefficient of the phase mismatch
\begin{equation}
    \Delta\phi_{GUI}\simeq4.7\nu^\frac{1}{5}.
    \label{GUI error near the ISCO} 
\end{equation}
\subsection{Phase Error in The Adiabatic Waveform} \label{sec:err_AD}
The error in phase of the adiabatic waveform is accumulated both at the ISCO's vicinity and at larger distances.

Similar to the GUI case, the  error accumulated near the ISCO is dominated by the mismatch between the adiabatic inspiral trajectory and the full dynamics near the ISCO, resulting in a phase error $\Delta\phi\sim\nu^{1/5}$.  
We evaluate the total phase mismatch between a test-particle following the adiabatic trajectory and one evolving under the full dynamics (Eq. \ref{equation of motion of the test-particle}) with a constant angular momentum loss rate, as given by Eq. (\ref{angular momentum diff}). 
In both cases, we assume that initially the particle is well outside the transition region, with an initial angular momentum given by Eq. (\ref{angular momentum diff}) with $\tau=10^5\cdot\nu^{-1/5}$.  
The resulting error in the adiabatic phase is:
\begin{equation}
    \Delta\phi_{ad}\simeq1.0\nu^\frac{1}{5}.
    \label{ADB error near the ISCO}
\end{equation}  
Combining the two results, Eqs. (\ref{GUI error near the ISCO}) and (\ref{ADB error near the ISCO}) 
give an upper limit for the mass-ratio of $\nu\lesssim10^{-4}$, above which our waveform accumulates more than a radian phase-shift around the ISCO.

It is well known that a ``transition regime" exists where the evolution is not accurately described neither by adiabatic evolution nor by a motion along a geodesic \citep{Ori_2000,buonanno_transition_2000}. Specifically, at the ISCO crossing, adiabatic evolution implies an infinite radial velocity, the GUI implies zero radial velocity, while in reality the radial velocity is of order $\nu^{3/5}$. 
Yet, our sharp connection accurately captures the phase and amplitude of the emitted GWs, including those emitted during this transition. 

At larger distances from the ISCO, the accumulated error is dominated by higher-order corrections to the adiabatic approximation \cite{wardell_gravitational_2023, Albertini_2022}. The post-adiabatic effects cause a radian phase shift at a separation an order of a unity beyond the ISCO. An exact calculation of the accumulated phase error of the adiabatic waveform requires a detailed treatment of the GW luminosity and binding energy beyond the first order in the mass-ratio, e.g., by using a GSF description \cite{wardell_gravitational_2023, Warburton_2021, Pound_2020, Le_Tiec_2012}.

We note that our procedure could be made more accurate -- at the expense of its universality and simplicity -- by incorporating the dynamics of the transition region to reduce the error accumulated around the ISCO and by adopting a post-adiabatic description for the GW emission beyond the ISCO, allowing the waveform to remain in phase at larger separations.

Finally, numerical errors in the calculation of the GW luminosity, $\delta P_{cir}$, introduce an additional source of phase error. As shown in Appendix \ref{app: ODE1}, these errors limit the validity of the adiabatic waveform to mass-ratios that satisfy $\nu\gg10\delta P_{cir}/P_{cir}$.
Our calculation estimates the GW luminosity with a relative error of $10^{-12}$, which is sufficient to accurately describe systems with $\nu$ as low as $10^{-10}$. This regime is relevant, for example, to mergers of brown dwarfs and SMBHs \citep[so called ``X-MRIs'';][]{Amaro_Seoane_2019,Aceves_2022}. 
\subsection{Comparison with the Literature}\label{sec: Comparison}
\begin{figure}
    \centering
        \includegraphics[width=\linewidth]{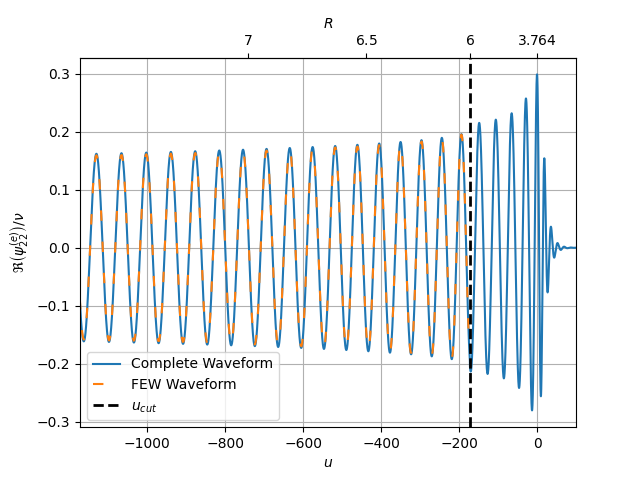}
        \caption{Comparison between the adiabatic waveform obtained by FEW (orange dashed line) \cite{chua_rapid_2021}, and our  approach (blue solid line), as described in Section \ref{sec:AD}. Our waveform includes both the adiabatic part and GUI part, while the FEW calculates only the adiabatic part. We present the real part of the dominant mode, $\psi_{22}^{(e)}$, assuming a mass-ratio $\nu=10^{-2}$. The black dashed line, located at $u_{cut}\approx-172$ (Eq. \ref{GUI cutting time}), marks the point of transition between the GUI waveform and the adiabatic waveform. The top axis shows the position of the test-particle. The complete waveform is shifted so it reaches its maximum at $u=0$, and the FEW is shifted so the two waveforms are aligned at $R=6.1$, the minimal separation that the FEW model follows.}
        
        \label{fig:FEW Waveform Comparison}
\end{figure}

\begin{figure}
    \centering
    \includegraphics[width=1\linewidth]{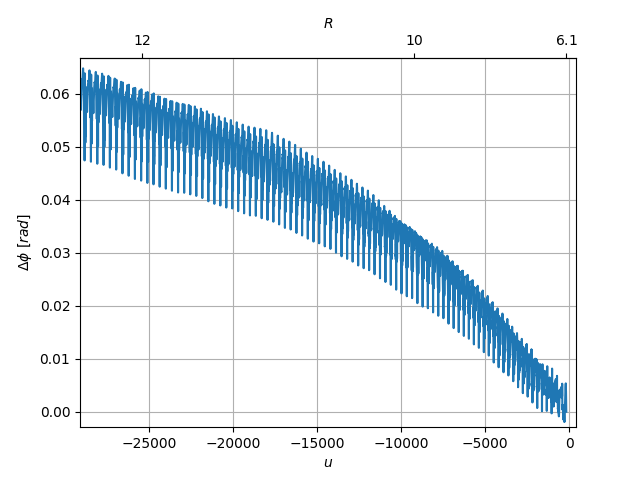}
    \caption{The phase mismatch in the dominant mode of our adiabatic waveform and the FEW \cite{chua_rapid_2021}, as a function of $u$. We present the results for mass-ratio $\nu=10^{-2}$. The waveforms are shifted to be aligned at $R=6.1$. The top axis shows the distance of the sBH from the ISCO. The oscillations originate from the FEW waveform.}
    \label{fig: FEW Phase Comparison}
\end{figure}

We compare the adiabatic part of our waveform with the FEW model \citep{chua_rapid_2021, Katz_2021}.
As expected, we find an excellent agreement between the two waveforms, with a total phase error below $0.1$ radian up to a separation of twice the ISCO.
Figure (\ref{fig:FEW Waveform Comparison}) shows both waveforms for a mass-ratio of $\nu=10^{-2}$, while Fig. (\ref{fig: FEW Phase Comparison}) presents their accumulated phase mismatch as a function of the retarded time $u$.

Notably, the current FEW package, as implemented in the BH perturbation toolkit \citep{BHPToolkit}, computes the GW luminosity with a relative error of $3\cdot10^{-7}$ \cite{chua_rapid_2021}, which introduces a non-negligible phase error for $\nu=10^{-6}$ along times of order $t\sim\nu^{-1}$ (see Appendix \ref{app: ODE1}). 
This results in a greater mismatch with our waveform, compared to its value at higher mass-ratios. 

We expect that this deviation will be resolved by refining the luminosity relative error threshold in the FEW code, bringing its accuracy closer to that of the \emph{Circular Orbit Self-force Data}\footnote{Both are part of the BH perturbation toolkit \citep{BHPToolkit}} (for more details see Appendix \ref{app: ODE}).


\begin{figure}[htbp]
    \begin{subfigure}{0.49\textwidth}
        \includegraphics[width=\linewidth]{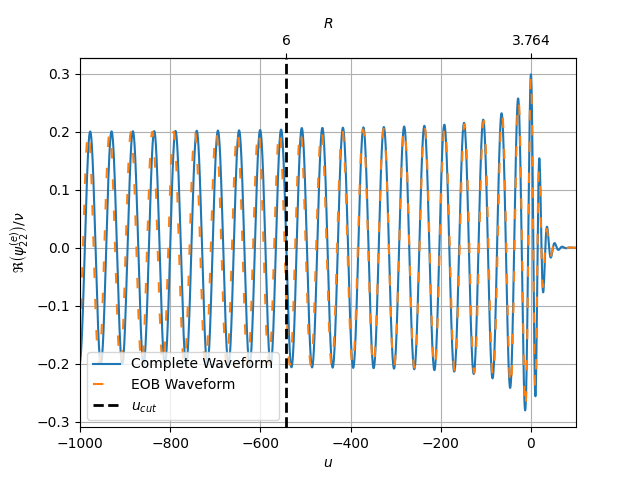}
        \centering
        \label{fig:EOB waveform comparison}
    \end{subfigure}
    \begin{subfigure}{0.49\textwidth}
        \centering
        \includegraphics[width=\linewidth]{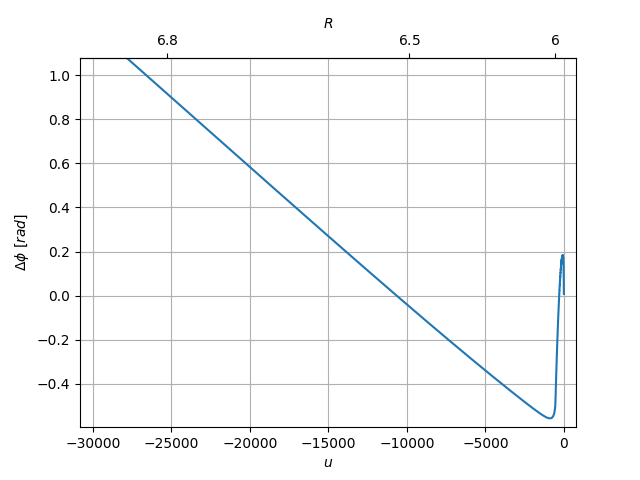}
        \label{fig:EOB phase comparison}
    \end{subfigure}
    \caption{{\bf Top panel:} Comparison between the EOB waveform \citep[dashed orange line;][]{nagar_time-domain_2018}, and our waveform (blue line). The black dashed line marks the transition between the GUI waveform and the adiabatic one, as given as given in Eq. (\ref{GUI cutting time}). We present the real part of the dominant mode, $\psi_{22}^{(e)}$, for a mass-ratio of $\nu=10^{-4}$. The waveforms are shifted so they both peak at $u=0$. 
    {\bf Bottom panel:} Phase mismatch between the two waveforms. In both figures, the top axis shows the position of the test-particle.}
    \label{figComparison to EOB Waveforms}
\end{figure}

Additionally, we compare our waveform to the TEOBResumS waveforms \cite{nagar_time-domain_2018}\footnote{We used the TEOBResumS version described by \cite{Albertini_2024}.}. Figure (\ref{figComparison to EOB Waveforms}) presents both waveforms for a mass-ratio $\nu=10^{-4}$. 
Evidently, the different methods closely agree along the plunge.
The mismatch reaches a radian phase shift around $R\approx6.8$ (corresponding to $u\simeq3\cdot10^4$), supporting our estimate (in Section \ref{sec:err_AD}) that the adiabatic waveforms stay in-phase up to a separation of an order of unity larger than the ISCO (see Section \ref{sec: Comparison} for further discussion),
yet further research is required to accurately quantify this error and the validity of the different methods in the extreme mass-ratio limit.

Furthermore, we compare the plunging time, defined as the interval between the ISCO crossing to the waveform peak, in the SEOBNRv5 model\footnote{In \cite{pompili2023layingfoundationeffectiveonebodywaveform}, this quantity is denoted as $\Delta t_{ISCO}^{22}$.} \cite{pompili2023layingfoundationeffectiveonebodywaveform} with the corresponding quantity in our model, $t_{plunge}$ (as given in Appendix \ref{app: Delta t}). 
In the SEOBNR approach, the plunging time is modeled using the ansatz
\begin{equation}
    t^{\ (SEOBNR)}_{plunge}=(a_0+a_1\nu+a_2\nu^2+a_3\nu^3)\nu^{-1/5+a_4\nu},
    \label{eq: SEOBNRv5 time diff}
\end{equation}
where the coefficients, given by Eq. (79) of \cite{pompili2023layingfoundationeffectiveonebodywaveform}, are calibrated using NR simulations, yielding high accuracy in the comparable to intermediate mass-ratio regime.
While in our model the ISCO crossing corresponds to the test-particle reaching $l_{ISCO}=\sqrt{12}$, in the SEOBNRv5 model it is defined with respect to the radial coordinate $R_{ISCO}=6$. As mentioned in section \ref{sec: Full}, these definitions are not equivalent \cite{Ori_2000,buonanno_transition_2000}, requiring an adjustment to our formula for a direct comparison. 
Evaluating the plunging time in our approach according to the SEOBNR definition, yields
\begin{equation}
    \begin{aligned}
        t_{plunge}=&80.312\nu^{-1/5}-112.7\\&+A\nu^{1/5}+O\left(\nu^{2/5}\right),
    \end{aligned}
    \label{eq: Adjusted Time Shift}
\end{equation}
where only the leading coefficient was changed due to the different ISCO crossing definition (compare Eq. \ref{eq: Adjusted Time Shift} with Eq. \ref{eq: plunge time approximation} in Appendix \ref{app: Delta t}).

Figure (\ref{fig: Time Diff Comparison}) shows the comparison between the SEOBNR plunging time (Eq. \ref{eq: SEOBNRv5 time diff}) and our result, Eq. (\ref{eq: Adjusted Time Shift}).
Despite their different functional form, the two estimates agree for intermediate mass-ratios, $10^{-2}\lesssim\nu\lesssim10^{-3}$.
However, as expected, they differ in both the extreme mass-ratio limit, $\nu\lesssim10^{-3}$, and the comparable-mass regime, $\nu\sim1$. 
In the extreme mass-ratio limit, our approach captures the dynamics more accurately, resulting in a different coefficient for the leading $\nu^{-1/5}$ term. In the comparable and intermediate mass-ratio regime, the SEOBNR model is more accurate, since it incorporates higher order corrections in the mass-ratio that are not included in our analysis.
\begin{figure}[htbp]
    \centering
    \includegraphics[width=0.9\linewidth]{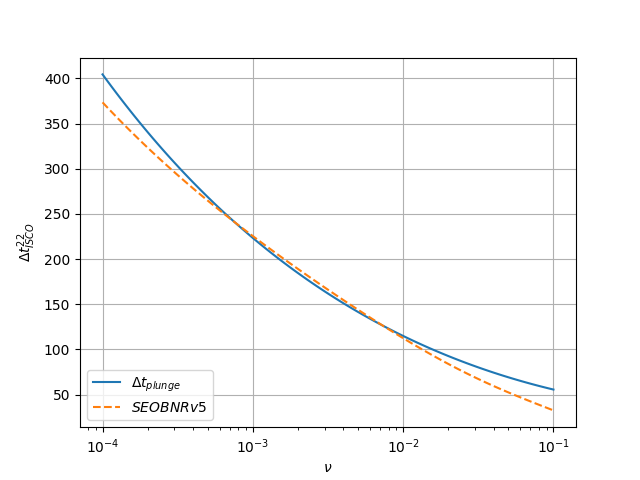}
    \caption{Comparison of the plunging time, from the ISCO crossing to the waveform peak in the SEOBNRv5 model (Eq. \ref{eq: SEOBNRv5 time diff}) to our estimation, Eq. (\ref{eq: Adjusted Time Shift}). The two results agree for intermediate mass-ratios, $10^{-2}\lesssim\nu\lesssim10^{-3}$. In the comparable mass-ratio regime ($\nu\rightarrow1$), the results differ due to higher order corrections in the mass-ratio, which are accounted for in the SEOBNR approach but not in our analysis. In the extreme mass-ratio regime, where our approach captures the dynamics more accurately, the deviation is caused by the difference in the coefficient of the leading $\nu^{-1/5}$ term.}
    \label{fig: Time Diff Comparison}
\end{figure}

\section{Discussion \& Summary}\label{sec: Conclusions}
We develop a simple approach to generate accurate universal EMRI waveforms. Utilizing the extreme mass-ratio, we analyze the dynamics in the test-particle limit. The evolution of the system can be divided into two parts: A slow adiabatic inspiral where the particle follows a quasi-circular orbit, and a rapid plunge toward the event horizon of the SMBH. 

For the adiabatic part, we generate the waveform by calculating the GW luminosity emitted by a particle on a circular orbit with a fractional precision of $10^{-12}$. The GUI waveform, which corresponds to the final plunge, is calculated by numerically solving the RWZ equation (based on our previous work \cite{rom_extreme_2022}). The two waveforms are sharply connected, according to the procedure described in Section {\ref{sec: Full}}. Our results are implemented in the \emph{Universal Waveforms} code \cite{UniversalWaveformsCode}.

We show that although the adiabatic approximation does not describe accurately the dynamics near the ISCO \cite{Ori_2000,buonanno_transition_2000}, it can be used to generate waveforms that remain on phase all the way from the ISCO up to distances of order unity beyond the ISCO. Thus, our simple approach allows us to generate waveforms that are accurate along timescales of order $t\sim\nu^{-1}$, containing both the long adiabatic part of the signal, and its peak. Furthermore, this waveform is universal, as it can be applied to any mass-ratio by renormalizing the time coordinate, amplitude and phase accordingly.

Nevertheless, our method could be extended by replacing adiabatic waveforms with post-adiabatic waveforms (such as the ones given by \cite{wardell_gravitational_2023}), potentially yielding a more accurate waveform that remains reliable through the merger and includes the peak of the GW signal. 
We compare our results to other models in the literature -- the FEW \cite{chua_rapid_2021, Katz_2021} and the TEOBResumS \cite{nagar_time-domain_2018, Albertini_2024} -- and show the mismatch between them.  We show an excellent match between our waveforms and the FEW waveforms for separations between twice the ISCO to the ISCO, and a less than one radian phase mismatch between our waveforms and the TEOBResumS waveforms up to a separation of order unit beyond the ISCO.

Finally, we note that our method can be generalized for spinning BHs and be extended to accurately describe larger separations by connecting the GUI waveform with post-adiabatic waveforms \citep[e.g.,][]{wardell_gravitational_2023}, rather than the simple adiabatic one used in this work.

\acknowledgments
We thank Scott Hughes and Rodrigo Panosso Macedo for useful discussions. This research was partially supported by an ISF grant, an NSF/BSF grant, an MOS grant and a GIF grant. B.R. acknowledges support from the Milner Foundation.
This work makes use of the Black Hole Perturbation Toolkit \cite{BHPToolkit}.
\appendix
\section{GW Luminosity Calculation }\label{app: ODE}
We compute the GW luminosity emitted by a test-particle on a circular orbit, $P_{cir}(r)$, by solving the RWZ equation  with outgoing wave boundary conditions \citep[For further details, see Section IV-A in][]{rom_extreme_2022}. 

For each mode, characterized by the multipolar indices $(l,m)$ and parity $\lambda$, we calculate the wave amplitudes at infinity and at the event horizon, denoted as $A_{lm}^{(\lambda)}(r)$ and $B_{lm}^{(\lambda)}(r)$, respectively. 
The GW luminosity is then given by \cite{nagar_gauge-invariant_2006,rom_extreme_2022}
\begin{equation}
\begin{aligned}
    P_{cir}(r)=\frac{1}{8\pi r^3}\sum_{\lambda lm}&m^2\frac{(l+2)!}{(l-2)!}\\
    &\times\left({A_{lm}^{(\lambda)}(r)}^2+{B_{lm}^{(\lambda)}(r)}^2\right).
    \label{GW luminosity}
\end{aligned}
\end{equation}

To accurately handle the boundary conditions at infinity and at the event horizon, we use the \emph{hyperboloidal method} \cite{Macedo_2022, Panosso_Macedo_2024}. Thus, the RWZ wave equation is reformulated in terms of the coordinate\footnote{
This transformation makes our formerly infinite grid finite, $r\in[2,\infty)\rightarrow\sigma\in[0,1]$.} $\sigma=2/r$ and the function $\bar{\phi}=e^{-sH(\sigma)}\psi_{lm}^{(\lambda)}$, where $s=-4im\Omega$ and 
\begin{equation}
    H(\sigma)=\frac{1}{2}\left(ln(1-\sigma)-\frac{1}{\sigma}+ln\sigma\right).
\end{equation}

The resulting equation is:
\begin{equation}
    \alpha_2(\sigma)\partial_\sigma^2\bar{\phi}+\alpha_1(\sigma)\partial_\sigma\Bar{\phi}+\alpha_0(\sigma)\bar{\phi}=0,
    \label{eq: hyperboloidal equation}
\end{equation}
where
\begin{equation}
    \begin{aligned}
        \alpha_0(\sigma)=&-s^2(1+\sigma)-2s\sigma\\
        &-\frac{4}{\sigma^2(1-\sigma)}V^{(\lambda)}_l\left(\frac{2}{\sigma}\right),\\
        \alpha_1(\sigma)=&s(1-2\sigma^2)-\sigma(3\sigma-2),\\
        \alpha_2(\sigma)=&\sigma^2(1-\sigma),
    \end{aligned}
    \label{eq: hyperboloidal coefficients}
\end{equation}
and $V^{(\lambda)}_l(r)$ is the curvature potential \cite{nagar_gauge-invariant_2006,rom_extreme_2022}.

Note that in this formulation, the outgoing wave boundary conditions are trivially satisfied\footnote{As long as $\bar{\phi}$ behaves regularly at the boundaries.} since $\alpha_2(\sigma)$ vanishes at the boundaries \cite{Macedo_2022, Panosso_Macedo_2024}.
Nonetheless, the vanishing of $\alpha_2(\sigma)$ at the boundaries introduces numerical difficulties at the boundaries vicinity. 
Therefore, we solve Eq. (\ref{eq: hyperboloidal equation}) analytically in the regions $\sigma\in[0,10^{-3}]$ and $\sigma\in[0.9,1]$. For the first region, where $\sigma\ll1$, we expand each function $\alpha_i(\sigma)$, given in Eq. (\ref{eq: hyperboloidal coefficients}), and $\bar{\phi}$ as a power series:
\begin{align}
\alpha_i(\sigma)&=\sum_{j=0}^\infty\alpha_{ij}\sigma^j,\\
\bar{\phi}(\sigma)&=\sum_{n=0}^\infty\bar{\phi}_n\sigma^n.
\end{align}
Similarly, in the second region, where $(1-\sigma)\ll1$, we expand these functions as power series in $(1-\sigma)$ instead of $\sigma$.
Substituting this to Eq. (\ref{eq: hyperboloidal equation}), we produce a recursion formula for the coefficients $\bar{\phi}_n$:
\begin{equation}
\begin{aligned}
\bar{\phi}_{N+1}=&-\frac{1}{\left(N(N+1)\alpha_{2,1}+(N+1)\alpha_{1,0}\right)}\sum_{n=0}^N C_n\bar{\phi}_n,
\end{aligned}
\end{equation}
where $C_n$ is given by
\begin{equation}
    C_n=n(n-1)\alpha_{2,N+2-n}+n\alpha_{1,N+1-n}+\alpha_{0,N-n}.
\end{equation}
We connect these analytical solutions near the boundaries with the numerical solution obtained for the rest of the grid, $\sigma\in[10^{-3},0.9]$.

\begin{figure}[bhtp]
    \centering
    \includegraphics[width=0.9\linewidth]{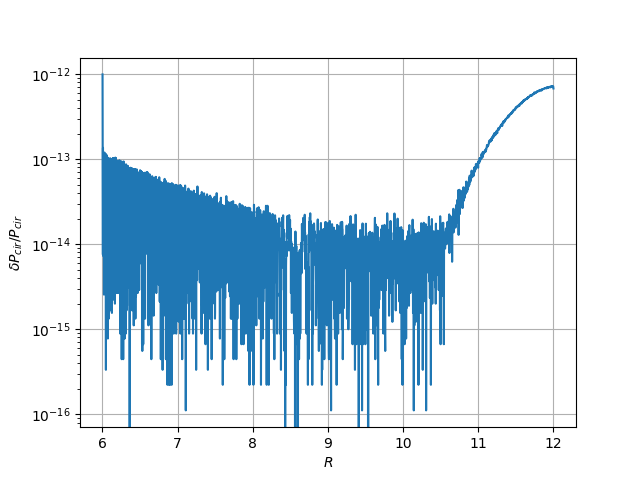}
    \caption{A comparison between the our interpolated GW luminosity and and the GW luminosity of \emph{Circular Orbit Self-force Data} in the BH perturbation toolkit \citep{BHPToolkit}. The relative error between the two stays below $10^{-12}$ for any separation below twice the ISCO, which is enough for the calculation of any EMRI waveform with any mass-ratio above $\nu\gtrsim10^{-10}$.}
    \label{fig: Circular Orbit Self-force Data Comparison}
\end{figure}

Finally, we calculate the contribution to the GW luminosity of all modes up to $l_{max}=30$, on a uniform grid between $R=6$ and $R=12$ with $\Delta R=0.002$, and use cubic spline interpolation for $P_{cir}(r)$ to solve Eqs. (\ref{eq:Adiabatic equations}). Fig. (\ref{fig: Circular Orbit Self-force Data Comparison}) shows a comparison between our interpolated GW luminosity and the GW luminosity of \emph{Circular Orbit Self-force Data} in the BH perturbation toolkit \citep{BHPToolkit}. The error stays below $10^{-12}$ for separations below twice the ISCO, which is enough for the calculation of any EMRI waveform with any mass-ratio below $\nu\lesssim10^{-10}$.
\subsection{Phase Error and Accuracy Requirement}\label{app: ODE1}
We estimate the phase error, $\delta\phi$, caused by numerical errors in the GW luminosity calculation during the adiabatic inspiral (Section \ref{sec:AD}). From the equations of motion, given in Eq. (\ref{eq:Adiabatic equations}), it follows that an average numerical error in the GW luminosity, $\delta\bar{P}_{cir}$,
leads to an accumulated error in the position of the test-particle that scales as
\begin{equation}
    \delta r\sim\frac{\Bar{\delta P_{cir}}}{P_{cir}} \left(r-6\right),
\end{equation}
assuming that initially (at $t=0$) the test-particle is at the ISCO.
This causes an accumulated error in phase of
\begin{equation}
    \delta\phi\sim\Omega\frac{\Bar{\delta P_{cir}}}{P_{cir}}|t|,
\end{equation}
where $\Omega$ is the orbital frequency. 

The relevant timescale to generate our adiabatic waveform from separations of twice the ISCO to the ISCO is $|t|\sim100/\nu$.
Therefore, the relative error must satisfy
\begin{equation}
    \frac{\delta\Bar{P}_{cir}}{P_{cir}}\ll 
    0.1\nu,
\end{equation}
where we substitute $\Omega\sim\Omega_{ISCO}\sim0.1$ (see Section \ref{sec: Full}).

In our numerical scheme, as described above, the GW luminosity is calculated up to a relative error of $10^{-12}$, allowing for an accurate calculation of the adiabatic waveform phase for mass-ratios of $\nu\gtrsim10^{-10}$.

\section{Calculation of the GUI Waveform Cutting Point}\label{app: Delta t}
We consider a point inside the ISCO, 
at a radius $r$ close to it, i.e., $(6-r)\ll1$, in a region where both the Ori-Thorne transition formalism \cite{Ori_2000} and the GUI approximation are applicable. 
We evaluate $t_{plunge}$ (defined in Section \ref{sec: Full}) as
\begin{equation}
    t_{plunge}=t_{OT}(r)+t_{GUI}(r),
    \label{t plunge as sum}
\end{equation}
where $t_{OT}(r)$ is the time it takes for the particle to inspiral in the transition regime, from the ISCO to the radius $r$, and $t_{GUI}(r)$ is the time it takes the particle to plunge along the GUI, from $r$ to $r_{peak}=3.764$. This radius corresponds to the peak of the GUI waveform. 

The expression for $t_{OT}(r)$ can be obtained by reversing and re-normalizing the expansion in Eq. (3.21) in ref. \cite{apte_exciting_2019}, yielding
\begin{equation}
\begin{aligned}
    t_{OT}(r)=
    &25.204P_{ISCO}^{-1/5}\nu^{-1/5}-\frac{3^{5/2}2^3}{\sqrt{6-r}}\\&+O\left(\nu(6-r)^{-3}\right).    
\end{aligned}
     \label{t_OT}
\end{equation}
Along the GUI, the time is given by $t_{GUI}(r)=g(r)-g(r_{peak})$, where 
\begin{equation}
\begin{aligned}
    g(r)=&\sqrt{\frac{8 r}{6-r}}\left(24-r\right)+44\sqrt{2}\sin^{-1}\left(\sqrt{\frac{6-r}{6}}\right)\\
    &-4\tanh^{-1}\left(\sqrt{\frac{6-r}{2r}}\right),
    \label{eq: g function}
\end{aligned}    
\end{equation}
as derived by \cite{Hadar_Kol_11,rom_extreme_2022}.
Expanding Eq. (\ref{eq: g function}) around $(6-r)\ll1$ then gives
\begin{equation}
    \begin{aligned}
t_{GUI}(r)=&\frac{3^{5/2}2^3}{\sqrt{6-r}}-g(r_{peak})
    +O\left(\sqrt{6-r}\right).      
    \end{aligned}
    \label{t_GUI}
\end{equation}
Substituting Eqs. (\ref{t_OT}) and (\ref{t_GUI}), using $r_{peak}=3.764$, to Eq. (\ref{t plunge as sum}) gives
\begin{equation}
\begin{aligned}
    t_{plunge}=& 101.6\nu^{-\frac{1}{5}}-112.7+A\nu^\frac{1}{5}+O\left(\nu^\frac{2}{5}\right).
\end{aligned}
    \label{eq: plunge time approximation}    
\end{equation}
Where the coefficient of the $O\left(\nu^{1/5}\right)$ term, marked as $A$, can be determined numerically,  
as given in Eq. (\ref{GUI cutting time}).

To determine the time at which we transition from the GUI waveform to the adiabatic one, denoted as $u_{cut}$, we add to $t_{plunge}$ (Eq. \ref{eq: plunge time approximation}) the additional signal travel time delay between the ISCO and $r_{peak}$, given by $\Delta r^*=r^*(6)-r^*(r_{peak})$. Thus, using Eq. (\ref{eq: plunge time approximation}) and the definition of the tortoise coordinate, we get
\begin{equation}
    u_{cut}=-101.6\nu^{-1/5}+108.8-A\nu^\frac{1}{5}+O\left(\nu^\frac{2}{5}\right).
    \label{eq: u cut} 
\end{equation}
Note that $u_{cut}$ is negative since we shifted the origin to be at the peak.

Finally, the accumulated phase along the GUI is given by \cite{hadar_post-isco_2011}
\begin{equation}
    \Phi(r)=\sqrt{\frac{12r}{6-r}},
\end{equation}
and it is used in our phase error estimation of the GUI waveform (see Section \ref{sec:err_GUI}).

\bibliographystyle{unsrt}
\bibliography{MyFirstPaperCitations}

\end{document}